\documentclass[twocolumn]{aastex631}

\usepackage[caption=false]{subfig} 
\usepackage[referable]{threeparttablex} 
\usepackage{booktabs} 

\begin{document}

\title{Ultracool Dwarfs Observed with the Spitzer Infrared Spectrograph: Equatorial Latitudes in L Dwarf Atmospheres are Cloudier}

\correspondingauthor{Genaro Su\'arez}
\email{gsuarez@amnh.org}

\author[0000-0002-2011-4924]{Genaro Su\'arez}
\affiliation{Department of Astrophysics, American Museum of Natural History, Central Park West at 79th Street, NY 10024, USA}

\author[0000-0003-0489-1528]{Johanna M. Vos}
\affiliation{School of Cosmic Physics, Dublin Institute for Advanced Studies, 31 Fitzwilliam Place, Dublin 2, D02 XF86, Ireland}
\affiliation{Department of Astrophysics, American Museum of Natural History, Central Park West at 79th Street, NY 10024, USA}

\author[0000-0003-3050-8203]{Stanimir Metchev}
\affiliation{Department of Physics and Astronomy, Western University, 1151 Richmond St, London, Ontario N6A 3K7, Canada}
\affiliation{Institute for Earth and Space Exploration, Western University, 1151 Richmond St, London, Ontario N6A 3K7, Canada}

\author[0000-0001-6251-0573]{Jacqueline K. Faherty}
\affiliation{Department of Astrophysics, American Museum of Natural History, Central Park West at 79th Street, NY 10024, USA}

\author[0000-0002-1821-0650]{Kelle Cruz}
\affiliation{Department of Astrophysics, American Museum of Natural History, Central Park West at 79th Street, NY 10024, USA}
\affiliation{Department of Physics and Astronomy, Hunter College, City University of New York, 695 Park Avenue, New York, NY 10065, USA}
\affiliation{Flatiron Institute, 162 Fifth Avenue, New York, NY 10010, USA}



\begin{abstract}
We report direct observational evidence for a latitudinal dependence of dust cloud opacity in ultracool dwarfs, indicating that equatorial latitudes are cloudier than polar latitudes. These results are based on a strong positive correlation between the viewing geometry and the mid-infrared silicate absorption strength in mid-L dwarfs using mid-infrared spectra from the Spitzer Space Telescope and spin axis inclination measurements from available information in the literature. We confirmed that the infrared color anomalies of L dwarfs positively correlate with dust cloud opacity and viewing geometry, where redder objects are inclined equator-on and exhibit more opaque dust clouds while dwarfs viewed at higher latitudes and with more transparent clouds are bluer. These results show the relevance of viewing geometry to explain the appearance of brown dwarfs and provide insight into the spectral diversity observed in substellar and planetary atmospheres. We also find a hint that dust clouds at similar latitudes may have higher opacity in low-surface gravity dwarfs than in higher-gravity objects.
\end{abstract}

\keywords{brown dwarfs --- stars: individual (2MASS~~J00361617+1821104, 2MASS~J03552337+1133437, 2MASS~J05012406$-$0010452, 2MASS~J11263991$-$5003550, 2MASS~J12560183$-$1257276, 2MASS~J14252798$-$3650229, 2MASS~J15074769$-$1627386, 2MASS~J17210390+3344160, 2MASS~J18212815+1414010, 2MASS~J21481633+4003594, 2MASS~J22443167+2043433) --- stars: atmospheres --- infrared: stars}


\section{Introduction}
\label{sec:introduction}

At temperatures $T_{\rm eff}\lesssim2000$~K the atmospheres of substellar objects are sufficiently cool to form clouds made of refractory oxides, silicates, and iron \citep{Lunine_etal1986,Fegley-Lodders1996,Tsuji_etal1996}. Silicate clouds are among the most prominent in the $\sim$2000--1000~K temperature range \citep[e.g.;][]{Burrows_etal1997,Ackerman-Marley2001}, roughly corresponding to effective temperatures of early-L to early-T dwarfs \citep{Filippazzo_etal2015}. Dust clouds play a critical role in shaping the emergent spectra of substellar atmospheres and, thus, understanding their effects is crucial to learn about the physics and chemistry of these atmospheres \citep[e.g.;][]{Knapp_etal2004,Burgasser_etal2008b,Faherty_etal2016}. 

The most easily observable spectroscopic feature that directly probes silicate clouds in ultracool (spectral types later than M7) atmospheres is a mid-infrared absorption between 9 and 13~$\mu$m \citep[e.g.;][]{Cushing_etal2006,Suarez-Metchev2022}. Prior to the James Webb Space Telescope (JWST), mid-infrared spectra of ultra-cool dwarfs were available only from the Infrared Spectrograph \citep[IRS; ][]{Houck_etal2004} on the Spitzer Space Telescope \citep{Werner_etal2004}. This allowed the first detection of the $\sim$10~$\mu$m silicate absorption in an L dwarf atmosphere \citep{Roellig_etal2004} followed by five more candidate or bonafide detections \citep{Cushing_etal2006,Looper_etal2008b}. Recently, \citet{Suarez-Metchev2022} reprocessed all Spitzer IRS spectra of ultracool dwarfs (late-M to T dwarfs) and presented dozens of spectroscopic silicate absorption detections in L dwarfs. Later and cooler objects could also exhibit silicate absorption. \citet{Vos_etal2023} inferred silicate clouds in two early-T dwarfs, similar to the clouds inferred by \citet{Burningham_etal2021} for an L dwarf with a strong silicate absorption.

Analyses of the mid-infrared silicate absorption have revealed that dust clouds in L-type atmospheres: $i)$ comprise a combination of $\lesssim$1~$\mu$m-sized amorphous silicate (enstatite, forsterite, pyroxene, and/or quartz) and iron grains \citep{Cushing_etal2008,Burningham_etal2021,Luna-Morley2021,Suarez-Metchev2023,Vos_etal2023}, $ii)$ form, then thicken, and sediment out of the visible atmospheres in the 2000 K $\lesssim T_{\rm{eff}} \lesssim 1300$ K range \citep{Suarez-Metchev2022}, $iii)$ are responsible for the color scatter \citep{Cushing_etal2006,Cushing_etal2008,Looper_etal2008b,Suarez-Metchev2022} and, most likely, for the observed variability due to cloud inhomogeneities \citep{Burgasser_etal2008b,Suarez-Metchev2022}, and $iv)$ exhibit gravity-dependent sedimentation properties, with dust clouds in low-gravity atmospheres dominated by heavier grains \citep{Suarez-Metchev2023}. For simplicity, we will refer to dust clouds in ultracool atmospheres as silicate clouds; however, as mentioned above, they could also include deeper iron layers \citep{Burningham_etal2021,Vos_etal2023}.

Dust cloud properties such as grain size, thickness, and patchiness---and effects attributed to these properties (e.g. variability amplitudes)---have been suggested to vary with latitude \citep[e.g.][]{Kirkpatrick_etal2010}. \citet{Vos_etal2017,Vos_etal2018,Vos_etal2020} observationally found that the near-infrared color anomaly and the variability amplitude of ultracool dwarfs positively correlate with their viewing inclination, where objects inclined equator-on appear redder and display higher variability amplitudes than near pole-on dwarfs. These relations are predicted by the atmospheric circulation models of brown dwarf and planetary atmospheres in \citet{Tan-Showman2021,Tan-Showman2021b}. The results are consistent with a picture in which cloud opacity is higher near the equator and lower close to the poles. Here we confirm  this scenario by directly analyzing the spectral signature of dust clouds.

Clouds are also an important component in the atmospheres of planets. They are common in the atmospheres of solar system's planets \citep[e.g.][]{Atreya_Wong2005,Gao_etal2021} and likely ubiquitous in extra-solar planet atmospheres \citep[e.g.;][]{Sing_etal2016,Crossfield-Kreidberg2017,Lothringer_etal2022}. Similar to clouds in substellar atmospheres, clouds in the  Jovian planets' atmospheres appear to be more opaque close to the equator and more transparent near the poles \citep{West_etal2009,Zhang_etal2013,Tollefson_etal2019}. These clouds influence most aspects of planetary atmospheres (e.g. radiation transport and atmospheric chemistry and dynamics) and, therefore, determine their emergent spectra \citep[e.g.;][]{Currie_etal2011,Marley_etal2013,Barstow_etal2017}.  A better understanding of the distribution of clouds in planetary and substellar atmospheres and their effects will allow us to explain the spectral diversity observed in brown dwarf and giant (extra-solar) planet atmospheres \citep[e.g.;][]{Looper_etal2008b,Burgasser_etal2008b,Faherty_etal2016,Liu_etal2016}.

We present in this study a strong and significant correlation between the opacity of dust clouds and their latitude in substellar atmospheres. In Section~\ref{sec:data} we describe the mid-infrared Spitzer spectra and the inclination measurements used in this work. Section~\ref{sec:results} presents the relations between silicate cloud opacity, viewing inclination, and infrared color anomaly. We discuss these results in Section~\ref{sec:discussion} and conclude in Section~\ref{sec:conclusions}.

\section{Data}
\label{sec:data}
To investigate how the opacity of silicate clouds may vary with latitude in a substellar atmosphere, we require information about viewing inclination angle and silicate cloud opacity. This information can inform how the spectral properties of brown dwarfs such as near-infrared colors relate to dust cloud properties or viewing geometry.

\subsection{Mid-infrared Spectroscopic Silicate Absorption}
In order to evaluate the silicate absorption feature, we use the Spitzer IRS mid-infrared spectroscopic sample of ultracool dwarfs in \citet{Suarez-Metchev2022}. We selected dwarfs with optical spectral types between L3 and L7 as they show a large diversity in silicate absorption strengths (see Figure~2a in \citealt{Suarez-Metchev2023}). We exclude L8 and later dwarfs since they may exhibit 7--9~$\mu$m methane absorption that affects the silicate absorption \citep{Suarez-Metchev2022}.

We measured the strength of the silicate absorption following the silicate index definition implemented in \citet{Suarez-Metchev2022} and improved in \citet[][see Figure 1 in that paper]{Suarez-Metchev2023}. Briefly, the index corresponds to the ratio of an interpolated continuum (using the 7.2--7.7~$\mu$m and 13.0--14.0~$\mu$m-wide windows) to the average flux at the center of the absorption (9.0--9.6~$\mu$m region).

The sample of L3--L7 dwarfs with Spitzer IRS spectra includes 39 objects with silicate index values roughly homogeneously distributed between 0.8 and 1.5. Silicate index values can be smaller than unity because the 7.2--7.7~$\mu$m continuum region contains residual water absorption in all L dwarfs, and methane absorption in late-L dwarfs, either of which leads to an underestimation of the actual continuum. Regardless, a small ($\lesssim$1.0) silicate index value means a weak or absent silicate absorption while a large ($\gtrsim$1.2) index value indicates a strong absorption.

\subsection{Viewing Geometry}
The viewing inclination angle (i) of an object can be determined from a combination of the following three parameters: $1)$ the projected rotational velocity ($v\ sin\ i$, where $v$ is the equatorial rotation velocity), obtained from the rotational broadening of spectral lines, $2)$ the rotation period ($P$), calculated from photometric variability, and $3)$ the radius ($R$), commonly obtained from spectral energy distribution (SED) fits combined with evolutionary models. Thus, the inclination angle is given by the ratio between the projected rotational velocity and the equatorial rotation velocity, as follows:

\begin{equation}
\label{eq:inclination}
    \textrm{sin}\ i = \frac{(v\ \textrm{sin}\ i)}{v} =  \frac{P}{2\pi R} (v\ \textrm{sin}\ i)
\end{equation}

We reviewed the literature for inclination measurements provided for dwarfs in our sample. As a starting point, we considered the viewing inclination angles in Table 3 by \citet{Vos_etal2020}, who combined their own spectroscopy-based determinations with data from the  literature. The \citeauthor{Vos_etal2020} compilation includes all brown dwarfs with measured variability at the time of publication. Twenty-three out of the 79 variable objects have inclination measurements in the compilation, most (20) of which are L dwarfs.

Nine of the 39 L3--L7 dwarfs with Spitzer IRS spectra have viewing angle measurements in \citet{Vos_etal2020}. In addition, we computed an inclination of $50\pm2$~deg for 0355+1133 using Equation~\ref{eq:inclination} and the Monte Carlo analysis outlined in \citet{Vos_etal2017,Vos_etal2020}, considering a rotation period of $P=9.53\pm0.19$~hr \citep{Vos_etal2022}, $v\ \textrm{sin}\ i = 12.31\pm0.15$~km~s$^{-1}$ \citep{Blake_etal2010}, and radius of $R=1.22\pm0.02~R_{\rm Jup}$ \citep{Vos_etal2022}. 

Since variability is a necessary property for inclination measurements, it is possible that variability in the SED may have an effect of the estimated radius used in the inclination calculation. However, \cite{Suarez_etal2021a} investigate this possibility for the variable object HN Peg B, finding that this effect is negligible for the observed amplitudes, so we do not consider this effect in this work.

In Table~\ref{tab:dwarfs} we list our final sample of 10 mid-L dwarfs with their optical and near-infrared spectral types, silicate indices, and inclination angles. An inclination of 90~deg means that the object is viewed equator-on and a value of 0~deg corresponds to a pole-on dwarf.

\subsection{Infrared Colors}
\label{sec:colors}
To explore how infrared colors of brown dwarfs may depend on the opacity of silicate clouds and their latitude in an atmosphere, we use near-infrared magnitudes from 2MASS \citep{Cutri_etal2003} and mid-infrared photometry from WISE \citep{Cutri_etal2013}. All {ten mid-L dwarfs in our sample have 2MASS and WISE W1 and W2 magnitudes, listed in Table~\ref{tab:phot}.

Before comparing the observed infrared colors of our sample to typical colors, we note that four (0355+1133, 0501$-$0010, 1425$-$3650, and 2244+2043) of the 10 selected mid-L dwarfs have $\gamma$ spectroscopic classification (Table~\ref{tab:dwarfs}) indicative of low surface gravity and therefore youth \citep{Cruz_etal2009,Gagne_etal2015,Faherty_etal2016,Manjavacas_etal2019,BardalezGagliuffi_etal2019}. Three (0355+1133, 1425$-$3650, and 2244+2043) of these low-gravity dwarfs are bonafide members of the 110--150~Myr-old \citep{Luhman_etal2005b,Barenfeld_etal2013} AB Doradus moving group \citep{Liu_etal2013,Gagne_etal2015,Vos_etal2018} and 0501$-$0010 is a probable member of the Columba or Carina moving groups \citep{Gagne_etal2015}, which are coeval at 20–40 Myr \citep{Torres_etal2008}. Conversely, the remaining six dwarfs with no spectroscopic signature of youth have $>$96\% probability of being field objects, according to the BANYAN $\Sigma $ algorithm \citep{Gagne_etal2018b}.

We defined the color anomaly for each selected dwarf as the difference between its observed infrared color and the average color (Table~\ref{tab:phot}) for its spectral type and surface gravity. For the four dwarfs with spectral and kinematic signatures of youth, we used the average colors of low-surface gravity L dwarfs in \citet{Faherty_etal2016}. Otherwise, we used average colors of field (non-low surface gravity, subdwarf, or young) L dwarfs from \citet{Faherty_etal2016}. Negative color anomaly values correspond to bluer than expected objects and positive values to redder than the average dwarfs.

\section{RESULTS: Positive correlations between silicate absorption strength, viewing inclination, and infrared color anomaly}

\label{sec:results}
We compare the viewing inclination angles, the silicate absorption strengths, and the color anomalies for the 10 L3--L7 dwarfs with Spitzer IRS spectra in Figure~\ref{fig:correlation}. The auxiliary color axis represents the color anomaly of each object based on 2MASS (left panel) or WISE (right panel) photometry. 

We observe a clear correlation between inclination and silicate index. Mid-L dwarfs with larger inclinations (viewed at lower latitudes) exhibit stronger silicate absorption, while mid-L dwarfs with lower inclinations (viewed at higher latitudes) have weaker or absent silicate absorption. The Pearson correlation coefficient $r$ is 0.9 with a $p$-value of 0.02\%, which is indicative of a strong, highly significant correlation \citep[e.g.;][]{Cohen1988}.

Figure~\ref{fig:correlation} also shows that both the 2MASS and WISE color anomalies positively correlate with inclination and silicate index: redder mid-L dwarfs have larger inclinations (more equator-on) and larger silicate indices than bluer objects. The Pearson correlation coefficients and $p$-values indicate that the relations between the silicate index and the $J-K_s$ or W1$-$W2 color anomalies are strong and significant ($0.8\leq r \leq 0.9$ and $p$-value$<$0.4\%). Similarly, the Pearson statistic is indicative of a strong and significant ($0.7\leq r \leq 0.8$ and $p$-value$<$3\%) correlation between the viewing inclination and the $J-K_s$ or W1$-$W2 color anomalies. These correlations between the $J-K_s$ color anomaly and the inclination or the silicate index are in agreement with the relations in \citet{Vos_etal2017} and \citet{Suarez-Metchev2022}, respectively.

\begin{figure*}
	\centering
	\subfloat{\includegraphics[width=.50\linewidth]{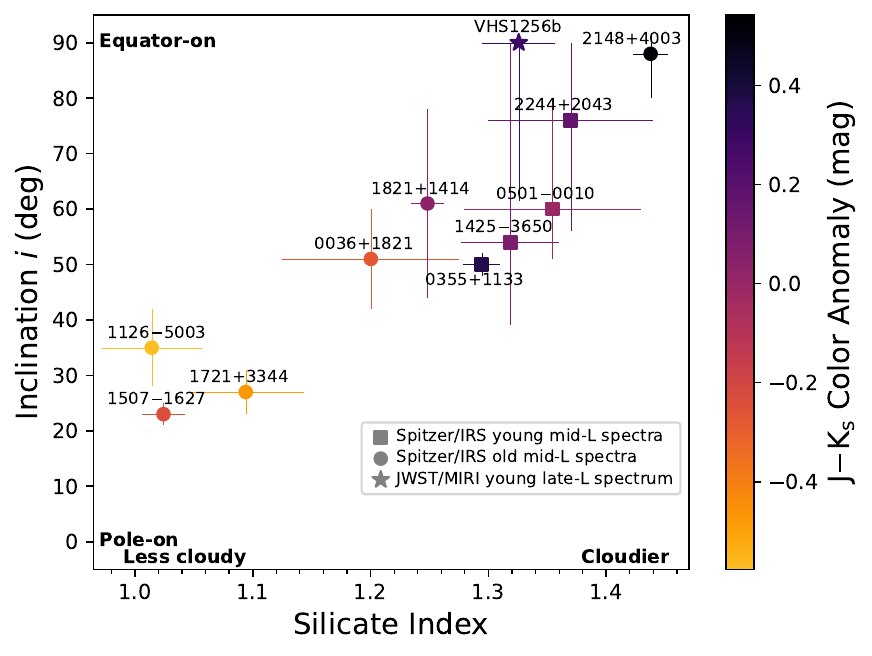}}
	\subfloat{\includegraphics[width=.50\linewidth]{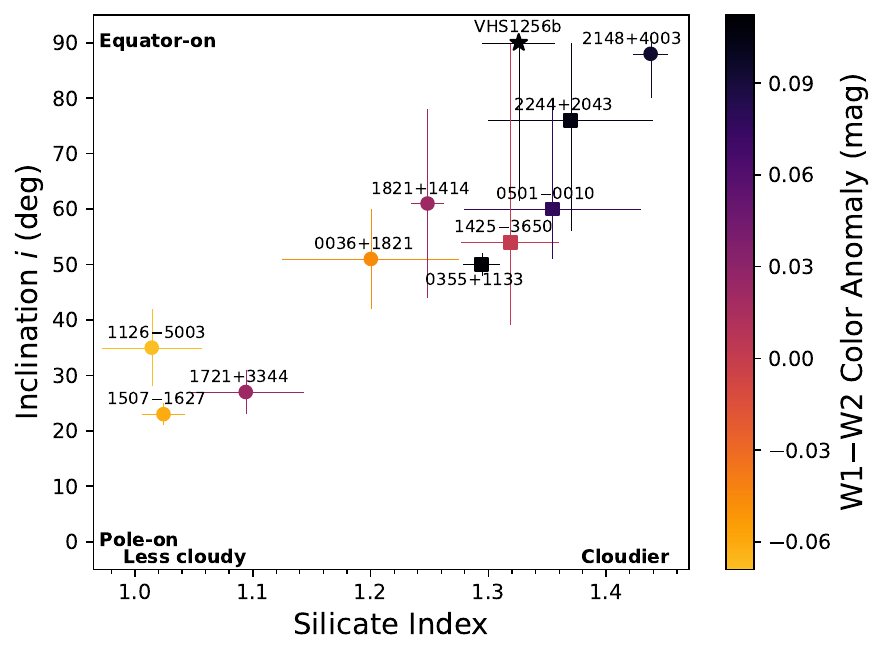}}
	\caption{Correlation between the spin axis inclination and the silicate index for 10 L3--L7 young (squares) and old (circles) dwarfs with Spitzer IRS spectra (Table~\ref{tab:dwarfs}). The auxiliary axis indicates the 2MASS $J-K_{\rm{s}}$ (left panel) and the WISE W1$-$W2 (right panel) color anomaly defined as the observed color of each object minus the average color (Table~\ref{tab:phot}) for its spectral type and surface gravity according to \citet{Faherty_etal2016}. The star symbol indicates the only object (VHS1256b) on the plot with a silicate index measurement not from Spitzer IRS spectra but from JWST MIRI (Secion~\ref{sec:VHS1256b}).}
	\label{fig:correlation}
\end{figure*}

\section{Discussion}
\label{sec:discussion}

\subsection{Latitudinal Dependence of Dust Cloud Opacity}
Literature studies have investigated the dependence of the infrared color anomaly of brown dwarfs with both viewing geometry and dust cloud opacity. \citet{Vos_etal2017} presented a positive correlation between the color anomaly and inclination of ultracool dwarfs. \citet{Suarez-Metchev2022} found that color anomaly is positively correlated with the opacity of silicate clouds in L dwarf atmospheres. This study connects both correlations. We report here for the first time direct observational evidence for a latitudinal dependence of dust cloud opacity in ultracool atmospheres. The observed positive correlation between viewing inclination and silicate absorption strength (Figure~\ref{fig:correlation}) indicates that the integrated optical depth of silicate clouds in L dwarf atmospheres changes with latitude, such that clouds are more opaque near the equator and more transparent closer to the poles. This is likely due to thicker clouds or denser cloud coverage at low latitudes and thinner clouds or sparser cloud coverage at higher latitudes.

Our observed equator-to-pole cloud opacity dependence in brown dwarf atmospheres agrees with observations of cloudier equatorial latitudes compared to pole regions in the atmospheres of the solar system gas giant planets \citep{West_etal2009,Zhang_etal2013,Tollefson_etal2019}. This is also consistent with the theoretical modeling for brown dwarf and planetary atmospheres by \citet{Tan-Showman2021b}, who predict that the vertical extent of dust clouds decreases from equator to pole due to the stronger effect of rotation at higher latitudes.

\subsection{The Appearance of Ultracool Atmospheres is Influenced by Viewing Geometry}
\label{sec:discussion_appearance}

The positive correlations between color anomaly and both inclination and silicate index (Figure~\ref{fig:correlation}) show that L dwarfs viewed equator-on appear cloudier and redder than similar spectral type dwarfs viewed pole-on. 
The color differences are more pronounced in the 2MASS $J-K_s$ near-infrared colors compared to the WISE W1$-$W2 mid-infrared colors. Figure~\ref{fig:correlation} shows that equator-on dwarfs can be redder than near pole-on dwarfs by more than a magnitude in $J-K_s$ and only about a fifth of magnitude in W1$-$W2. Such a significantly stronger $J-K_s$ color difference may be due to the fact that this color probes a larger (by a factor of three to four) pressure range than the W1$-$W2 color \citep{Burningham_etal2021}. These color differences are comparable to the color scatter observed in L dwarfs with the same subtype \citep[e.g.;][]{Cushing_etal2008,Faherty_etal2016}.

Considering reported evidence that dust cloud properties are potentially responsible for various photometric phenomena in L-type atmospheres, such as a large color scatter \citep[e.g.;][]{Knapp_etal2004,Cushing_etal2008,Faherty_etal2016,Suarez-Metchev2022}, ubiquitous variability \citep[e.g.;][]{Burgasser_etal2002,Marley_etal2012,Radigan_etal2014,Metchev_etal2015,Suarez-Metchev2022, Vos_etal2022}, and a range of variability amplitudes \citep{Vos_etal2020}, the correlation presented here between viewing inclination and silicate cloud opacity suggests that the appearance of brown dwarfs is strongly influenced by their viewing geometry. This provides new insight into the diversity of observed properties in brown dwarf and planetary atmospheres \citep[e.g.;][]{Faherty_etal2009,Faherty_etal2016,Liu_etal2016,Schneider_etal2023}.

The latitudinal dependence of dust cloud optical depth should also produce an equator-to-pole brightness temperature gradation of the photosphere. Lower cloud opacity at the poles would reveal deeper, hotter atmospheric layers, while greater cloud opacity at the equator would reveal cooler, lower-density upper layers. Such latitudinal temperature differences have been reported from observations of the Jovian planets \citep{Ingersoll_1990,Tollefson_etal2019} and 
predicted with atmospheric dynamics models of brown dwarfs and giant planets by \citet{Showman_Kaspi2013} and \citet{Tan-Showman2021b}. 

The higher dust cloud opacity near equatorial regions might also explain the higher near-infrared variability amplitudes observed in red and near equator-on dwarfs by \citet{Vos_etal2017,Vos_etal2020}. The atmospheric circulation models of \citet{Tan-Showman2021b} predict that large-scale cloud structures increase in size towards the equator. The combination of greater cloud opacity, as inferred here, and larger cloud structures toward the equator could explain the larger-amplitude variability of objects seen at higher inclinations.

\subsection{Extension of the Spitzer Results to the L/T Transition with JWST}
\label{sec:VHS1256b}

\citet{Miles_etal2023} presented the first JWST MIRI ($R\sim2000$) spectrum of a substellar mass object showing the $\sim$10~$\mu$m silicate absorption for the L-to-T transition planetary-mass companion VHS J125060.192$-$125723.9b (VHS1256b). VHS1256b is a young object \citep[140$\pm$20~Myr;][]{Dupuy_etal2023} seen equator-on \citep[$90_{-28}^{+0}$~deg; ][]{Zhou_etal2020}, and is substantially redder than expected ($J-K_{\rm s}$ and W1$-$W2 color anomalies of 0.294 mag and 0.196 mag, respectively). Based on the correlations in Figure~\ref{fig:correlation}, this object should exhibit strong silicate absorption. We confirmed this by measuring a high silicate index (1.33$\pm$0.03) after convolving its MIRI spectrum to the $R\sim100$ IRS resolution. Hence, VHS1256b is in good agreement with the inferred latitudinal dependence of dust cloud opacity in mid-L dwarfs, and extends the result to later-type objects (see star symbol in Figure~\ref{fig:correlation}). 

We note, however, that for late-L objects such as VHS1256b \citep[L8/L7 optical/near-infrared types;][]{Gauza_etal2015} the silicate index definition in \citet{Suarez-Metchev2022,Suarez-Metchev2023} may no longer adequately represent the strength of silicate absorption. This is because the short-wavelength continuum window could be underestimated due to the onset of the 7.65~$\mu$m methane absorption \citep{Suarez-Metchev2022}, causing an underestimation of the silicate index. We observed in Figure 11 of \citet{Miles_etal2023} that it might be the case for the VHS1256b MIRI spectrum, where the 7.2--7.7~$\mu$m silicate absorption continuum window appears depressed compared to the IRS spectrum of an earlier dwarf (L4.5; 2224-0158) with no evidence of methane. Although we show the position of VHS1256b in Figure~\ref{fig:correlation}, this object was not considered to evaluate the significance of the trends between inclination, silicate index, and near-infrared color anomaly in Section~\ref{sec:results}.

\subsection{Similar Latitudes May Be Cloudier in Young Atmospheres}
\label{sec:young_cloudier_atmospheres}

A closer inspection of Figure~\ref{fig:correlation} suggests that dust cloud opacity may also depend on surface gravity. Three (0355+1133, 0501$-$0010, and 1425$-$3650) of the five dwarfs with inclinations $\approx$50--60~deg have both a systematically larger silicate index and low-surface gravity (Section~\ref{sec:colors}; $\log g<4.5$ in \citealt{Gagne_etal2015} and \citealt{Filippazzo_etal2015}), although the silicate indices for the five dwarfs are consistent within the uncertainties. The other two dwarfs (0036+1821 and 1821+1414) in this inclination range and with lower silicate indices have evidence of high-surface gravity (field gravity classification in \citealt{Martin_etal2017} and $\log g>5.1$ in \citealt{Filippazzo_etal2015}) and are 99.9\% probable field dwarfs (Section~\ref{sec:colors}). This may suggest that dust clouds at similar latitudes are thicker or denser in low-surface gravity L dwarfs than in higher gravity objects with similar spectral types. 

There is a low-surface gravity object (VHS1256b) and a high-gravity (2148+4003) dwarf both viewed close to equator-on which could also inform on the cloud opacity-surface gravity trend at equatorial latitudes. These objects apparently follow an opposite trend to the one discussed above, with the young object having a lower silicate index. Nevertheless, as we discussed in Section \ref{sec:VHS1256b}, the silicate index of VHS1256b may be underestimated due to the appearance of methane absorption at the short-wavelength continuum region as a consequence of its later (L8) spectral type. Additionally, being at the L/T spectral type transition, VHS1256b may be experiencing the sedimentation of silicate clouds and, therefore, exhibit less clouds in its mid-infrared photosphere compared to earlier type dwarfs viewed equator-on. Consequently, the VHS1256b silicate index is not directly comparable to the indices of L3--L7 dwarfs and likely is not representative of the surface gravity dependence of their cloudiness.

A higher cloud optical depth in younger L-type atmospheres, as suggested by the subsample of dwarfs with $\approx$50--60~deg inclinations, may be due to lower sedimentation efficiencies of dust clouds compared to the cloud settling in higher gravity atmospheres \citep[e.g.; ][]{Ackerman-Marley2001,Stephens_etal2009,Faherty_etal2012,Faherty_etal2016,Suarez-Metchev2023}. Such dustier atmospheres, if patchy, could explain the higher variability amplitudes observed in low-surface gravity dwarfs \citep{Metchev_etal2015,Vos_etal2022} due to a higher contrast between cloudy and less cloudy regions, and the higher variability occurrence \citep{Metchev_etal2015,Vos_etal2019,Vos_etal2022} compared to  those in higher gravity objects. Thus, variability amplitudes might be enhanced in cloudy young dwarf atmospheres, which is a similar effect to the maximized variability amplitudes in closer to equator-on dwarfs  discussed in Section~\ref{sec:discussion_appearance}.

This tentative result about a cloud opacity difference between low and high-surface gravity dwarfs with a similar viewing geometry is encouraging for future JWST mid-infrared spectroscopic observations of young and old L dwarfs. Higher SNR and resolution mid-infrared L-type spectra will also enable robust modeling of the silicate absorption to further evaluate how cloud properties such as grain size, chemistry, and vertical extent may change with latitude.

\section{Conclusions}
\label{sec:conclusions}

We present a strong and significant positive correlation between the optical depth of silicate clouds in L dwarf atmospheres and their viewing inclinations using mid-infrared spectra from Spitzer IRS and spin axis inclination measurements. This represents the first direct observational evidence of a latitudinal dependence of integrated dust cloud opacity, with the equatorial regions appearing cloudier than the polar regions. This is consistent with what has been observed in the atmospheres of the Jovian planets \citep[e.g.;][]{Tollefson_etal2019}.

We also show that dust cloud opacity and viewing inclination are correlated with infrared color anomalies of L dwarfs. That is, objects viewed closer to equator-on and exhibiting stronger silicate absorption are redder than dwarfs viewed closer to pole-on and exhibiting lower cloud opacity, as reported before \citep{Vos_etal2017,Suarez-Metchev2022}.

The results suggest that the observed spectral diversity in brown dwarfs is at least partially explained by their viewing geometry.

There is also a hint that dust clouds at similar latitudes may be more opaque in young atmospheres than in old atmospheres. This is encouraging for JWST mid-infrared spectroscopic observations of low- and high-gravity dwarfs.

\section*{Acknowledgements}
G. S. and J. K. F. acknowledge support from NSF award \#2009177. 
J. M. V. acknowledges support from a Royal Society - Science Foundation Ireland University Research Fellowship (URF$\backslash$1$\backslash$221932). S.\ M.\ acknowledges support from the Canadian Space Agency under the Flights
and Fieldwork for the Advancement of Science and Technology
program (21FAUWOB12).

\facility{Spitzer}

\defcitealias{BardalezGagliuffi_etal2019}{Bard19}
\defcitealias{Burgasser_etal2008b}{Burg08}
\defcitealias{Cruz_etal2009}{Cruz09}
\defcitealias{Cruz_etal2003}{Cruz03}
\defcitealias{Gagne_etal2015}{Gagn15}
\defcitealias{Gauza_etal2015}{Gauz15}
\defcitealias{Kirkpatrick_etal2000}{Kirk00}
\defcitealias{Kirkpatrick_etal2008}{Kirk08}
\defcitealias{Kirkpatrick_etal2010}{Kirk10}
\defcitealias{Knapp_etal2004}{Knap04}
\defcitealias{Looper_etal2008b}{Loop08}
\defcitealias{Reid_etal2008}{Reid08}
\defcitealias{Vos_etal2017}{Vos17}
\defcitealias{Vos_etal2018}{Vos18}
\defcitealias{Vos_etal2020}{Vos20}
\defcitealias{Zhou_etal2020}{Zhou20}

\begin{table*}
\begin{flushleft}
\caption{Mid-L dwarfs with mid-infrared spectra and viewing inclination measurements.}
  \small
  \label{tab:dwarfs}
\begin{tabular}{lcccccccc}
    \toprule
	2MASS Name                    & Short Name   & SpT$_{\textrm{opt}}$ & Ref.                              & SpT$_{\textrm{IR}}$ & Ref.                                    & Silicate Index$^{\rm a}$ & Inclination      & Ref.                      \\ 
		                          &              &                      &                                   &                     &                                         &                & (deg)            &                           \\ 
    \midrule
	J00361617$+$1821104           & 0036$+$1821  & L3.5                 & \citetalias{Kirkpatrick_etal2000} &  L4                 & \citetalias{Knapp_etal2004}             & 1.20$\pm$0.08  & 51$\pm$9         & \citetalias{Vos_etal2017} \\ 
	J03552337$+$1133437$^{\rm b}$ & 0355$+$1133  & L5$\gamma$           & \citetalias{Cruz_etal2009}        &  L4$\gamma$         & \citetalias{BardalezGagliuffi_etal2019} & 1.29$\pm$0.02  & 50$\pm$2         & This study                \\ 
	J05012406$-$0010452$^{\rm b}$ & 0501$-$0010  & L4$\gamma$           & \citetalias{Cruz_etal2009}        &  L4$\gamma$         & \citetalias{Gagne_etal2015}             & 1.35$\pm$0.08  & 60$_{-9}^{+20}$  & \citetalias{Vos_etal2020} \\ 
	J11263991$-$5003550           & 1126$-$5003  & L4.5                 & \citetalias{Burgasser_etal2008b}  &  L6.5p              & \citetalias{Burgasser_etal2008b}        & 1.01$\pm$0.04  & 35$\pm$7         & \citetalias{Vos_etal2020} \\ 
	J14252798$-$3650229$^{\rm b}$ & 1425$-$3650  & L3                   & \citetalias{Reid_etal2008}        &  L4$\gamma$         & \citetalias{Gagne_etal2015}             & 1.32$\pm$0.04  & 54$_{-15}^{+36}$ & \citetalias{Vos_etal2020} \\ 
	J15074769$-$1627386           & 1507$-$1627  & L5                   & \citetalias{Kirkpatrick_etal2000} &  L5.5               & \citetalias{Knapp_etal2004}             & 1.02$\pm$0.02  & 23$\pm$2         & \citetalias{Vos_etal2017} \\ 
	J17210390$+$3344160           & 1721$+$3344  & L3                   & \citetalias{Cruz_etal2003}        &  L5                 & \citetalias{Burgasser_etal2008b}        & 1.09$\pm$0.05  & 27$\pm$4         & \citetalias{Vos_etal2017} \\ 
	J18212815$+$1414010           & 1821$+$1414  & L4.5                 & \citetalias{Looper_etal2008b}     &  L5p                & \citetalias{Kirkpatrick_etal2010}       & 1.25$\pm$0.01  & 61$\pm$17        & \citetalias{Vos_etal2017} \\ 
	J21481633$+$4003594           & 2148$+$4003  & L6                   & \citetalias{Looper_etal2008b}     &  L6.5p              & \citetalias{Kirkpatrick_etal2010}       & 1.44$\pm$0.01  & 88$_{-8}^{+2}$   & \citetalias{Vos_etal2017} \\ 
	J22443167$+$2043433$^{\rm b}$ & 2244$+$2043  & L6.5                 & \citetalias{Kirkpatrick_etal2008} &  L6--L8$\gamma$     & \citetalias{Gagne_etal2015}             & 1.37$\pm$0.07  & 76$_{-20}^{+14}$ & \citetalias{Vos_etal2018} \\ 
	\bottomrule
 	\end{tabular}
	\par $^{\rm a}$Silicate index measurements using Spitzer/IRS spectra \citep{Suarez-Metchev2022}.\\
    \par $^{\rm b}$Objects with spectral signatures of low-surface gravity \citep[$\gamma$ classification; ][]{Cruz_etal2009} and kinematic memberships to known young moving groups (Section~\ref{sec:colors}).
    \par \textbf{References.} Bard19: \citet{BardalezGagliuffi_etal2019}, Burg08: \citet{Burgasser_etal2008b}, Cruz09: \citet{Cruz_etal2009}, Cruz03: \citet{Cruz_etal2003}, Gagn15: \citet{Gagne_etal2015}, Gauz15: \citet{Gauza_etal2015}, Kirk00: \citet{Kirkpatrick_etal2000}, Kirk08: \citet{Kirkpatrick_etal2008}, Kirk10: \citet{Kirkpatrick_etal2010}, Knap04: \citet{Knapp_etal2004}, Loop08: \citet{Looper_etal2008b}, Reid08: \citet{Reid_etal2008}, Vos17: \citet{Vos_etal2017}, Vos18: \citet{Vos_etal2018}, Vos20: \citet{Vos_etal2020}, Zhou20: \citet{Zhou_etal2020}.\\
\end{flushleft}
\end{table*}

\begin{table*}
\begin{flushleft}
\caption{Near-infrared 2MASS and mid-infrared WISE photometry  of the L3--L7 dwarfs in this study together with average infrared colors for each object.}   \small
  \label{tab:phot}
	\begin{tabular}{cccccccc}
    \toprule
	Short Name   & $J$              & $H$              & $K_{\rm s}$      & W1               & W2               & $(J-K_{\rm s})_{\rm mean}^a$ & (W1$-$W2)$_{\rm mean}^a$ \\
	             & (mag)            & (mag)            & (mag)            & (mag)            & (mag)            & (mag)                        & (mag)                    \\
    \midrule                                                                                                                                                                %
	0036$+$1821  & 12.466$\pm$0.027 & 11.588$\pm$0.030 & 11.058$\pm$0.021 & 10.516$\pm$0.024 & 10.237$\pm$0.020 & 1.675                        & 0.325                    \\
	0355$+$1133  & 14.050$\pm$0.024 & 12.530$\pm$0.031 & 11.526$\pm$0.021 & 10.528$\pm$0.023 &  9.943$\pm$0.021 & 2.153                        & 0.473                    \\
	0501$-$0010  & 14.982$\pm$0.038 & 13.713$\pm$0.034 & 12.963$\pm$0.035 & 12.050$\pm$0.024 & 11.518$\pm$0.022 & 2.020                        & 0.455                    \\
	1126$-$5003  & 13.997$\pm$0.032 & 13.284$\pm$0.036 & 12.829$\pm$0.029 & 12.218$\pm$0.024 & 11.942$\pm$0.022 & 1.745                        & 0.345                    \\
	1425$-$3650  & 13.747$\pm$0.028 & 12.575$\pm$0.022 & 11.805$\pm$0.027 & 10.998$\pm$0.022 & 10.576$\pm$0.020 & 1.850                        & 0.421                    \\
	1507$-$1627  & 12.830$\pm$0.027 & 11.895$\pm$0.024 & 11.312$\pm$0.026 & 10.667$\pm$0.023 & 10.378$\pm$0.022 & 1.750                        & 0.350                    \\
	1721$+$3344  & 13.625$\pm$0.023 & 12.952$\pm$0.026 & 12.489$\pm$0.020 & 11.933$\pm$0.024 & 11.601$\pm$0.022 & 1.610                        & 0.310                    \\
	1821$+$1414  & 13.431$\pm$0.024 & 12.396$\pm$0.019 & 11.650$\pm$0.021 & 10.845$\pm$0.029 & 10.478$\pm$0.021 & 1.745                        & 0.345                    \\
	2148$+$4003  & 14.147$\pm$0.029 & 12.783$\pm$0.030 & 11.765$\pm$0.023 & 10.739$\pm$0.023 & 10.235$\pm$0.021 & 1.840                        & 0.410                    \\
	2244$+$2043  & 16.476$\pm$0.140 & 14.999$\pm$0.066 & 14.022$\pm$0.073 & 12.777$\pm$0.024 & 12.108$\pm$0.024 & 2.290                        & 0.564                    \\
    \bottomrule
 	\end{tabular}
	\par $^a$Average colors for the spectral type and (low or high) surface gravity of each object according to \citet{Faherty_etal2016}.
\end{flushleft}
\end{table*}

\bibliography{mybib_Suarez}

\begin{thebibliography}{}
\expandafter\ifx\csname natexlab\endcsname\relax\def\natexlab#1{#1}\fi
\providecommand{\url}[1]{\href{#1}{#1}}
\providecommand{\dodoi}[1]{doi:~\href{http://doi.org/#1}{\nolinkurl{#1}}}
\providecommand{\doeprint}[1]{\href{http://ascl.net/#1}{\nolinkurl{http://ascl.net/#1}}}
\providecommand{\doarXiv}[1]{\href{https://arxiv.org/abs/#1}{\nolinkurl{https://arxiv.org/abs/#1}}}

\bibitem[{{Ackerman} \& {Marley}(2001)}]{Ackerman-Marley2001}
{Ackerman}, A.~S., \& {Marley}, M.~S. 2001, \apj, 556, 872,
  \dodoi{10.1086/321540}

\bibitem[{{Atreya} \& {Wong}(2005)}]{Atreya_Wong2005}
{Atreya}, S.~K., \& {Wong}, A.-S. 2005, \ssr, 116, 121,
  \dodoi{10.1007/s11214-005-1951-5}

\bibitem[{{Bardalez Gagliuffi} {et~al.}(2019){Bardalez Gagliuffi}, {Burgasser},
  {Schmidt}, {Theissen}, {Gagn{\'e}}, {Gillon}, {Sahlmann}, {Faherty},
  {Gelino}, {Cruz}, {Skrzypek}, \& {Looper}}]{BardalezGagliuffi_etal2019}
{Bardalez Gagliuffi}, D.~C., {Burgasser}, A.~J., {Schmidt}, S.~J., {et~al.}
  2019, \apj, 883, 205, \dodoi{10.3847/1538-4357/ab253d}

\bibitem[{{Barenfeld} {et~al.}(2013){Barenfeld}, {Bubar}, {Mamajek}, \&
  {Young}}]{Barenfeld_etal2013}
{Barenfeld}, S.~A., {Bubar}, E.~J., {Mamajek}, E.~E., \& {Young}, P.~A. 2013,
  \apj, 766, 6, \dodoi{10.1088/0004-637X/766/1/6}

\bibitem[{{Barstow} {et~al.}(2017){Barstow}, {Aigrain}, {Irwin}, \&
  {Sing}}]{Barstow_etal2017}
{Barstow}, J.~K., {Aigrain}, S., {Irwin}, P.~G.~J., \& {Sing}, D.~K. 2017,
  \apj, 834, 50, \dodoi{10.3847/1538-4357/834/1/50}

\bibitem[{{Blake} {et~al.}(2010){Blake}, {Charbonneau}, \&
  {White}}]{Blake_etal2010}
{Blake}, C.~H., {Charbonneau}, D., \& {White}, R.~J. 2010, \apj, 723, 684,
  \dodoi{10.1088/0004-637X/723/1/684}

\bibitem[{{Burgasser} {et~al.}(2008){Burgasser}, {Looper}, {Kirkpatrick},
  {Cruz}, \& {Swift}}]{Burgasser_etal2008b}
{Burgasser}, A.~J., {Looper}, D.~L., {Kirkpatrick}, J.~D., {Cruz}, K.~L., \&
  {Swift}, B.~J. 2008, \apj, 674, 451, \dodoi{10.1086/524726}

\bibitem[{{Burgasser} {et~al.}(2002){Burgasser}, {Kirkpatrick}, {Brown},
  {Reid}, {Burrows}, {Liebert}, {Matthews}, {Gizis}, {Dahn}, {Monet}, {Cutri},
  \& {Skrutskie}}]{Burgasser_etal2002}
{Burgasser}, A.~J., {Kirkpatrick}, J.~D., {Brown}, M.~E., {et~al.} 2002, \apj,
  564, 421, \dodoi{10.1086/324033}

\bibitem[{{Burningham} {et~al.}(2021){Burningham}, {Faherty}, {Gonzales},
  {Marley}, {Visscher}, {Lupu}, {Gaarn}, {Fabienne Bieger}, {Freedman}, \&
  {Saumon}}]{Burningham_etal2021}
{Burningham}, B., {Faherty}, J.~K., {Gonzales}, E.~C., {et~al.} 2021, \mnras,
  506, 1944, \dodoi{10.1093/mnras/stab1361}

\bibitem[{{Burrows} {et~al.}(1997){Burrows}, {Marley}, {Hubbard}, {Lunine},
  {Guillot}, {Saumon}, {Freedman}, {Sudarsky}, \& {Sharp}}]{Burrows_etal1997}
{Burrows}, A., {Marley}, M., {Hubbard}, W.~B., {et~al.} 1997, \apj, 491, 856,
  \dodoi{10.1086/305002}

\bibitem[{Cohen(1988)}]{Cohen1988}
Cohen, J. 1988, {Statistical Power Analysis for the Behavioral Sciences
  (Hillsdale, NJ: L. Erlbaum Associates)}

\bibitem[{{Crossfield} \& {Kreidberg}(2017)}]{Crossfield-Kreidberg2017}
{Crossfield}, I. J.~M., \& {Kreidberg}, L. 2017, \aj, 154, 261,
  \dodoi{10.3847/1538-3881/aa9279}

\bibitem[{{Cruz} {et~al.}(2009){Cruz}, {Kirkpatrick}, \&
  {Burgasser}}]{Cruz_etal2009}
{Cruz}, K.~L., {Kirkpatrick}, J.~D., \& {Burgasser}, A.~J. 2009, \aj, 137,
  3345, \dodoi{10.1088/0004-6256/137/2/3345}

\bibitem[{{Cruz} {et~al.}(2003){Cruz}, {Reid}, {Liebert}, {Kirkpatrick}, \&
  {Lowrance}}]{Cruz_etal2003}
{Cruz}, K.~L., {Reid}, I.~N., {Liebert}, J., {Kirkpatrick}, J.~D., \&
  {Lowrance}, P.~J. 2003, \aj, 126, 2421, \dodoi{10.1086/378607}

\bibitem[{{Currie} {et~al.}(2011){Currie}, {Burrows}, {Itoh}, {Matsumura},
  {Fukagawa}, {Apai}, {Madhusudhan}, {Hinz}, {Rodigas}, {Kasper}, {Pyo}, \&
  {Ogino}}]{Currie_etal2011}
{Currie}, T., {Burrows}, A., {Itoh}, Y., {et~al.} 2011, \apj, 729, 128,
  \dodoi{10.1088/0004-637X/729/2/128}

\bibitem[{{Cushing} {et~al.}(2006){Cushing}, {Roellig}, {Marley}, {Saumon},
  {Leggett}, {Kirkpatrick}, {Wilson}, {Sloan}, {Mainzer}, {Van Cleve}, \&
  {Houck}}]{Cushing_etal2006}
{Cushing}, M.~C., {Roellig}, T.~L., {Marley}, M.~S., {et~al.} 2006, \apj, 648,
  614, \dodoi{10.1086/505637}

\bibitem[{{Cushing} {et~al.}(2008){Cushing}, {Marley}, {Saumon}, {Kelly},
  {Vacca}, {Rayner}, {Freedman}, {Lodders}, \& {Roellig}}]{Cushing_etal2008}
{Cushing}, M.~C., {Marley}, M.~S., {Saumon}, D., {et~al.} 2008, \apj, 678,
  1372, \dodoi{10.1086/526489}

\bibitem[{{Cutri} {et~al.}(2003){Cutri}, {Skrutskie}, {van Dyk}, {Beichman},
  {Carpenter}, {Chester}, {Cambresy}, {Evans}, {Fowler}, {Gizis}, {Howard},
  {Huchra}, {Jarrett}, {Kopan}, {Kirkpatrick}, {Light}, {Marsh}, {McCallon},
  {Schneider}, {Stiening}, {Sykes}, {Weinberg}, {Wheaton}, {Wheelock}, \&
  {Zacarias}}]{Cutri_etal2003}
{Cutri}, R.~M., {Skrutskie}, M.~F., {van Dyk}, S., {et~al.} 2003, {2MASS All
  Sky Catalog of point sources.}

\bibitem[{{Cutri} {et~al.}(2013){Cutri}, {Wright}, {Conrow}, {Fowler},
  {Eisenhardt}, {Grillmair}, {Kirkpatrick}, {Masci}, {McCallon}, {Wheelock},
  {Fajardo-Acosta}, {Yan}, {Benford}, {Harbut}, {Jarrett}, {Lake}, {Leisawitz},
  {Ressler}, {Stanford}, {Tsai}, {Liu}, {Helou}, {Mainzer}, {Gettngs},
  {Gonzalez}, {Hoffman}, {Marsh}, {Padgett}, {Skrutskie}, {Beck}, {Papin}, \&
  {Wittman}}]{Cutri_etal2013}
{Cutri}, R.~M., {Wright}, E.~L., {Conrow}, T., {et~al.} 2013, yCat, 2328, 0

\bibitem[{{Dupuy} {et~al.}(2023){Dupuy}, {Liu}, {Evans}, {Best}, {Pearce},
  {Sanghi}, {Phillips}, \& {Bardalez Gagliuffi}}]{Dupuy_etal2023}
{Dupuy}, T.~J., {Liu}, M.~C., {Evans}, E.~L., {et~al.} 2023, \mnras, 519, 1688,
  \dodoi{10.1093/mnras/stac3557}

\bibitem[{{Faherty} {et~al.}(2009){Faherty}, {Burgasser}, {Cruz}, {Shara},
  {Walter}, \& {Gelino}}]{Faherty_etal2009}
{Faherty}, J.~K., {Burgasser}, A.~J., {Cruz}, K.~L., {et~al.} 2009, \aj, 137,
  1, \dodoi{10.1088/0004-6256/137/1/1}

\bibitem[{{Faherty} {et~al.}(2012){Faherty}, {Burgasser}, {Walter}, {Van der
  Bliek}, {Shara}, {Cruz}, {West}, {Vrba}, \&
  {Anglada-Escud{\'e}}}]{Faherty_etal2012}
{Faherty}, J.~K., {Burgasser}, A.~J., {Walter}, F.~M., {et~al.} 2012, \apj,
  752, 56, \dodoi{10.1088/0004-637X/752/1/56}

\bibitem[{{Faherty} {et~al.}(2016){Faherty}, {Riedel}, {Cruz}, {Gagne},
  {Filippazzo}, {Lambrides}, {Fica}, {Weinberger}, {Thorstensen}, {Tinney},
  {Baldassare}, {Lemonier}, \& {Rice}}]{Faherty_etal2016}
{Faherty}, J.~K., {Riedel}, A.~R., {Cruz}, K.~L., {et~al.} 2016, \apjs, 225,
  10, \dodoi{10.3847/0067-0049/225/1/10}

\bibitem[{{Fegley} \& {Lodders}(1996)}]{Fegley-Lodders1996}
{Fegley}, Bruce, J., \& {Lodders}, K. 1996, \apjl, 472, L37,
  \dodoi{10.1086/310356}

\bibitem[{{Filippazzo} {et~al.}(2015){Filippazzo}, {Rice}, {Faherty}, {Cruz},
  {Van Gordon}, \& {Looper}}]{Filippazzo_etal2015}
{Filippazzo}, J.~C., {Rice}, E.~L., {Faherty}, J., {et~al.} 2015, \apj, 810,
  158, \dodoi{10.1088/0004-637X/810/2/158}

\bibitem[{{Gagn{\'e}} {et~al.}(2015){Gagn{\'e}}, {Faherty}, {Cruz},
  {Lafreni{\'e}re}, {Doyon}, {Malo}, {Burgasser}, {Naud}, {Artigau},
  {Bouchard}, {Gizis}, \& {Albert}}]{Gagne_etal2015}
{Gagn{\'e}}, J., {Faherty}, J.~K., {Cruz}, K.~L., {et~al.} 2015, \apjs, 219,
  33, \dodoi{10.1088/0067-0049/219/2/33}

\bibitem[{{Gagn{\'e}} {et~al.}(2018){Gagn{\'e}}, {Mamajek}, {Malo}, {Riedel},
  {Rodriguez}, {Lafreni{\`e}re}, {Faherty}, {Roy-Loubier}, {Pueyo}, {Robin}, \&
  {Doyon}}]{Gagne_etal2018b}
{Gagn{\'e}}, J., {Mamajek}, E.~E., {Malo}, L., {et~al.} 2018, \apj, 856, 23,
  \dodoi{10.3847/1538-4357/aaae09}

\bibitem[{{Gao} {et~al.}(2021){Gao}, {Wakeford}, {Moran}, \&
  {Parmentier}}]{Gao_etal2021}
{Gao}, P., {Wakeford}, H.~R., {Moran}, S.~E., \& {Parmentier}, V. 2021, Journal
  of Geophysical Research (Planets), 126, e06655, \dodoi{10.1029/2020JE006655}

\bibitem[{{Gauza} {et~al.}(2015){Gauza}, {B{\'e}jar}, {P{\'e}rez-Garrido},
  {Zapatero Osorio}, {Lodieu}, {Rebolo}, {Pall{\'e}}, \&
  {Nowak}}]{Gauza_etal2015}
{Gauza}, B., {B{\'e}jar}, V. J.~S., {P{\'e}rez-Garrido}, A., {et~al.} 2015,
  \apj, 804, 96, \dodoi{10.1088/0004-637X/804/2/96}

\bibitem[{{Houck} {et~al.}(2004){Houck}, {Roellig}, {van Cleve}, {Forrest},
  {Herter}, {Lawrence}, {Matthews}, {Reitsema}, {Soifer}, {Watson}, {Weedman},
  {Huisjen}, {Troeltzsch}, {Barry}, {Bernard-Salas}, {Blacken}, {Brandl},
  {Charmandaris}, {Devost}, {Gull}, {Hall}, {Henderson}, {Higdon}, {Pirger},
  {Schoenwald}, {Sloan}, {Uchida}, {Appleton}, {Armus}, {Burgdorf},
  {Fajardo-Acosta}, {Grillmair}, {Ingalls}, {Morris}, \&
  {Teplitz}}]{Houck_etal2004}
{Houck}, J.~R., {Roellig}, T.~L., {van Cleve}, J., {et~al.} 2004, \apjs, 154,
  18, \dodoi{10.1086/423134}

\bibitem[{{Ingersoll}(1990)}]{Ingersoll_1990}
{Ingersoll}, A.~P. 1990, Science, 248, 308,
  \dodoi{10.1126/science.248.4953.308}

\bibitem[{{Kirkpatrick} {et~al.}(2000){Kirkpatrick}, {Reid}, {Liebert},
  {Gizis}, {Burgasser}, {Monet}, {Dahn}, {Nelson}, \&
  {Williams}}]{Kirkpatrick_etal2000}
{Kirkpatrick}, J.~D., {Reid}, I.~N., {Liebert}, J., {et~al.} 2000, \aj, 120,
  447, \dodoi{10.1086/301427}

\bibitem[{{Kirkpatrick} {et~al.}(2008){Kirkpatrick}, {Cruz}, {Barman},
  {Burgasser}, {Looper}, {Tinney}, {Gelino}, {Lowrance}, {Liebert},
  {Carpenter}, {Hillenbrand}, \& {Stauffer}}]{Kirkpatrick_etal2008}
{Kirkpatrick}, J.~D., {Cruz}, K.~L., {Barman}, T.~S., {et~al.} 2008, \apj, 689,
  1295, \dodoi{10.1086/592768}

\bibitem[{{Kirkpatrick} {et~al.}(2010){Kirkpatrick}, {Looper}, {Burgasser},
  {Schurr}, {Cutri}, {Cushing}, {Cruz}, {Sweet}, {Knapp}, {Barman},
  {Bochanski}, {Roellig}, {McLean}, {McGovern}, \&
  {Rice}}]{Kirkpatrick_etal2010}
{Kirkpatrick}, J.~D., {Looper}, D.~L., {Burgasser}, A.~J., {et~al.} 2010,
  \apjs, 190, 100, \dodoi{10.1088/0067-0049/190/1/100}

\bibitem[{{Knapp} {et~al.}(2004){Knapp}, {Leggett}, {Fan}, {Marley}, {Geballe},
  {Golimowski}, {Finkbeiner}, {Gunn}, {Hennawi}, {Ivezi{\'c}}, {Lupton},
  {Schlegel}, {Strauss}, {Tsvetanov}, {Chiu}, {Hoversten}, {Glazebrook},
  {Zheng}, {Hendrickson}, {Williams}, {Uomoto}, {Vrba}, {Henden}, {Luginbuhl},
  {Guetter}, {Munn}, {Canzian}, {Schneider}, \& {Brinkmann}}]{Knapp_etal2004}
{Knapp}, G.~R., {Leggett}, S.~K., {Fan}, X., {et~al.} 2004, \aj, 127, 3553,
  \dodoi{10.1086/420707}

\bibitem[{{Liu} {et~al.}(2013){Liu}, {Dupuy}, \& {Allers}}]{Liu_etal2013}
{Liu}, M.~C., {Dupuy}, T.~J., \& {Allers}, K.~N. 2013, Astronomische
  Nachrichten, 334, 85, \dodoi{10.1002/asna.201211783}

\bibitem[{{Liu} {et~al.}(2016){Liu}, {Dupuy}, \& {Allers}}]{Liu_etal2016}
---. 2016, \apj, 833, 96, \dodoi{10.3847/1538-4357/833/1/96}

\bibitem[{{Looper} {et~al.}(2008){Looper}, {Kirkpatrick}, {Cutri}, {Barman},
  {Burgasser}, {Cushing}, {Roellig}, {McGovern}, {McLean}, {Rice}, {Swift}, \&
  {Schurr}}]{Looper_etal2008b}
{Looper}, D.~L., {Kirkpatrick}, J.~D., {Cutri}, R.~M., {et~al.} 2008, \apj,
  686, 528, \dodoi{10.1086/591025}

\bibitem[{Lothringer {et~al.}(2022)Lothringer, Sing, Rustamkulov, Wakeford,
  Stevenson, Nikolov, Lavvas, Spake, \& Winch}]{Lothringer_etal2022}
Lothringer, J.~D., Sing, D.~K., Rustamkulov, Z., {et~al.} 2022, Nature
  Astronomy, 604, 49, \dodoi{10.1038/s41586-022-04453-2}

\bibitem[{{Luhman} {et~al.}(2005){Luhman}, {Stauffer}, \&
  {Mamajek}}]{Luhman_etal2005b}
{Luhman}, K.~L., {Stauffer}, J.~R., \& {Mamajek}, E.~E. 2005, \apjl, 628, L69,
  \dodoi{10.1086/432617}

\bibitem[{{Luna} \& {Morley}(2021)}]{Luna-Morley2021}
{Luna}, J.~L., \& {Morley}, C.~V. 2021, \apj, 920, 146,
  \dodoi{10.3847/1538-4357/ac1865}

\bibitem[{{Lunine} {et~al.}(1986){Lunine}, {Hubbard}, \&
  {Marley}}]{Lunine_etal1986}
{Lunine}, J.~I., {Hubbard}, W.~B., \& {Marley}, M.~S. 1986, \apj, 310, 238,
  \dodoi{10.1086/164678}

\bibitem[{{Manjavacas} {et~al.}(2019){Manjavacas}, {Apai}, {Zhou}, {Lew},
  {Schneider}, {Metchev}, {Miles-P{\'a}ez}, {Radigan}, {Marley}, {Cowan},
  {Karalidi}, {Burgasser}, {Bedin}, {Lowrance}, \&
  {Kauffmann}}]{Manjavacas_etal2019}
{Manjavacas}, E., {Apai}, D., {Zhou}, Y., {et~al.} 2019, \aj, 157, 101,
  \dodoi{10.3847/1538-3881/aaf88f}

\bibitem[{{Marley} {et~al.}(2013){Marley}, {Ackerman}, {Cuzzi}, \&
  {Kitzmann}}]{Marley_etal2013}
{Marley}, M.~S., {Ackerman}, A.~S., {Cuzzi}, J.~N., \& {Kitzmann}, D. 2013, in
  Comparative Climatology of Terrestrial Planets, ed. S.~J. {Mackwell}, A.~A.
  {Simon-Miller}, J.~W. {Harder}, \& M.~A. {Bullock}, 367--392,
  \dodoi{10.2458/azu_uapress_9780816530595-ch015}

\bibitem[{{Marley} {et~al.}(2012){Marley}, {Saumon}, {Cushing}, {Ackerman},
  {Fortney}, \& {Freedman}}]{Marley_etal2012}
{Marley}, M.~S., {Saumon}, D., {Cushing}, M., {et~al.} 2012, \apj, 754, 135,
  \dodoi{10.1088/0004-637X/754/2/135}

\bibitem[{{Martin} {et~al.}(2017){Martin}, {Mace}, {McLean}, {Logsdon}, {Rice},
  {Kirkpatrick}, {Burgasser}, {McGovern}, \& {Prato}}]{Martin_etal2017}
{Martin}, E.~C., {Mace}, G.~N., {McLean}, I.~S., {et~al.} 2017, \apj, 838, 73,
  \dodoi{10.3847/1538-4357/aa6338}

\bibitem[{{Metchev} {et~al.}(2015){Metchev}, {Heinze}, {Apai}, {Flateau},
  {Radigan}, {Burgasser}, {Marley}, {Artigau}, {Plavchan}, \&
  {Goldman}}]{Metchev_etal2015}
{Metchev}, S.~A., {Heinze}, A., {Apai}, D., {et~al.} 2015, \apj, 799, 154,
  \dodoi{10.1088/0004-637X/799/2/154}

\bibitem[{{Miles} {et~al.}(2023){Miles}, {Biller}, {Patapis}, {Worthen},
  {Rickman}, {Hoch}, {Skemer}, {Perrin}, {Whiteford}, {Chen}, {Sargent},
  {Mukherjee}, {Morley}, {Moran}, {Bonnefoy}, {Petrus}, {Carter}, {Choquet},
  {Hinkley}, {Ward-Duong}, {Leisenring}, {Millar-Blanchaer}, {Pueyo}, {Ray},
  {Sallum}, {Stapelfeldt}, {Stone}, {Wang}, {Absil}, {Balmer}, {Boccaletti},
  {Bonavita}, {Booth}, {Bowler}, {Chauvin}, {Christiaens}, {Currie},
  {Danielski}, {Fortney}, {Girard}, {Grady}, {Greenbaum}, {Henning}, {Hines},
  {Janson}, {Kalas}, {Kammerer}, {Kennedy}, {Kenworthy}, {Kervella}, {Lagage},
  {Lew}, {Liu}, {Macintosh}, {Marino}, {Marley}, {Marois}, {Matthews},
  {Matthews}, {Mawet}, {McElwain}, {Metchev}, {Meyer}, {Molliere}, {Pantin},
  {Quirrenbach}, {Rebollido}, {Ren}, {Schneider}, {Vasist}, {Wyatt}, {Zhou},
  {Briesemeister}, {Bryan}, {Calissendorff}, {Cantalloube}, {Cugno}, {De
  Furio}, {Dupuy}, {Factor}, {Faherty}, {Fitzgerald}, {Franson}, {Gonzales},
  {Hood}, {Howe}, {Kraus}, {Kuzuhara}, {Lagrange}, {Lawson}, {Lazzoni}, {Liu},
  {Llop-Sayson}, {Lloyd}, {Martinez}, {Mazoyer}, {Quanz}, {Redai}, {Samland},
  {Schlieder}, {Tamura}, {Tan}, {Uyama}, {Vigan}, {Vos}, {Wagner}, {Wolff},
  {Ygouf}, {Zhang}, {Zhang}, \& {Zhang}}]{Miles_etal2023}
{Miles}, B.~E., {Biller}, B.~A., {Patapis}, P., {et~al.} 2023, \apjl, 946, L6,
  \dodoi{10.3847/2041-8213/acb04a}

\bibitem[{{Radigan} {et~al.}(2014){Radigan}, {Lafreni{\`e}re}, {Jayawardhana},
  \& {Artigau}}]{Radigan_etal2014}
{Radigan}, J., {Lafreni{\`e}re}, D., {Jayawardhana}, R., \& {Artigau}, E. 2014,
  \apj, 793, 75, \dodoi{10.1088/0004-637X/793/2/75}

\bibitem[{{Reid} {et~al.}(2008){Reid}, {Cruz}, {Kirkpatrick}, {Allen},
  {Mungall}, {Liebert}, {Lowrance}, \& {Sweet}}]{Reid_etal2008}
{Reid}, I.~N., {Cruz}, K.~L., {Kirkpatrick}, J.~D., {et~al.} 2008, \aj, 136,
  1290, \dodoi{10.1088/0004-6256/136/3/1290}

\bibitem[{{Roellig} {et~al.}(2004){Roellig}, {Van Cleve}, {Sloan}, {Wilson},
  {Saumon}, {Leggett}, {Marley}, {Cushing}, {Kirkpatrick}, {Mainzer}, \&
  {Houck}}]{Roellig_etal2004}
{Roellig}, T.~L., {Van Cleve}, J.~E., {Sloan}, G.~C., {et~al.} 2004, \apjs,
  154, 418

\bibitem[{{Schneider} {et~al.}(2023){Schneider}, {Burgasser}, {Bruursema},
  {Munn}, {Vrba}, {Caselden}, {Kabatnik}, {Rothermich}, {Sainio}, {Bickle},
  {Dahm}, {Meisner}, {Kirkpatrick}, {Su{\'a}rez}, {Gagn{\'e}}, {Faherty},
  {Vos}, {Kuchner}, {Williams}, {Bardalez Gagliuffi}, {Aganze}, {Hsu},
  {Theissen}, {Cushing}, {Marocco}, {Casewell}, \& {Backyard Worlds: Planet 9
  Collaboration}}]{Schneider_etal2023}
{Schneider}, A.~C., {Burgasser}, A.~J., {Bruursema}, J., {et~al.} 2023, \apjl,
  943, L16, \dodoi{10.3847/2041-8213/acb0cd}

\bibitem[{{Showman} \& {Kaspi}(2013)}]{Showman_Kaspi2013}
{Showman}, A.~P., \& {Kaspi}, Y. 2013, \apj, 776, 85,
  \dodoi{10.1088/0004-637X/776/2/85}

\bibitem[{{Sing} {et~al.}(2016){Sing}, {Fortney}, {Nikolov}, {Wakeford},
  {Kataria}, {Evans}, {Aigrain}, {Ballester}, {Burrows}, {Deming},
  {D{\'e}sert}, {Gibson}, {Henry}, {Huitson}, {Knutson}, {Lecavelier Des
  Etangs}, {Pont}, {Showman}, {Vidal-Madjar}, {Williamson}, \&
  {Wilson}}]{Sing_etal2016}
{Sing}, D.~K., {Fortney}, J.~J., {Nikolov}, N., {et~al.} 2016, \nat, 529, 59,
  \dodoi{10.1038/nature16068}

\bibitem[{{Stephens} {et~al.}(2009){Stephens}, {Leggett}, {Cushing}, {Marley},
  {Saumon}, {Geballe}, {Golimowski}, {Fan}, \& {Noll}}]{Stephens_etal2009}
{Stephens}, D.~C., {Leggett}, S.~K., {Cushing}, M.~C., {et~al.} 2009, \apj,
  702, 154, \dodoi{10.1088/0004-637X/702/1/154}

\bibitem[{{Su{\'a}rez} \& {Metchev}(2022)}]{Suarez-Metchev2022}
{Su{\'a}rez}, G., \& {Metchev}, S. 2022, \mnras, 513, 5701,
  \dodoi{10.1093/mnras/stac1205}

\bibitem[{{Su{\'a}rez} \& {Metchev}(2023)}]{Suarez-Metchev2023}
---. 2023, arXiv e-prints, arXiv:2306.01119, \dodoi{10.48550/arXiv.2306.01119}

\bibitem[{{Su{\'a}rez} {et~al.}(2021){Su{\'a}rez}, {Metchev}, {Leggett},
  {Saumon}, \& {Marley}}]{Suarez_etal2021a}
{Su{\'a}rez}, G., {Metchev}, S., {Leggett}, S.~K., {Saumon}, D., \& {Marley},
  M.~S. 2021, \apj, 920, 99, \dodoi{10.3847/1538-4357/ac1418}

\bibitem[{{Tan} \& {Showman}(2021{\natexlab{a}})}]{Tan-Showman2021}
{Tan}, X., \& {Showman}, A.~P. 2021{\natexlab{a}}, \mnras, 502, 678,
  \dodoi{10.1093/mnras/stab060}

\bibitem[{{Tan} \& {Showman}(2021{\natexlab{b}})}]{Tan-Showman2021b}
---. 2021{\natexlab{b}}, \mnras, 502, 2198, \dodoi{10.1093/mnras/stab097}

\bibitem[{{Tollefson} {et~al.}(2019){Tollefson}, {de Pater}, {Luszcz-Cook}, \&
  {DeBoer}}]{Tollefson_etal2019}
{Tollefson}, J., {de Pater}, I., {Luszcz-Cook}, S., \& {DeBoer}, D. 2019, \aj,
  157, 251, \dodoi{10.3847/1538-3881/ab1fdf}

\bibitem[{{Torres} {et~al.}(2008){Torres}, {Quast}, {Melo}, \&
  {Sterzik}}]{Torres_etal2008}
{Torres}, C.~A.~O., {Quast}, G.~R., {Melo}, C.~H.~F., \& {Sterzik}, M.~F. 2008,
  in Handbook of Star Forming Regions, Volume II, ed. B.~{Reipurth}, Vol.~5,
  757

\bibitem[{{Tsuji} {et~al.}(1996){Tsuji}, {Ohnaka}, \& {Aoki}}]{Tsuji_etal1996}
{Tsuji}, T., {Ohnaka}, K., \& {Aoki}, W. 1996, \aap, 305, L1

\bibitem[{{Vos} {et~al.}(2017){Vos}, {Allers}, \& {Biller}}]{Vos_etal2017}
{Vos}, J.~M., {Allers}, K.~N., \& {Biller}, B.~A. 2017, \apj, 842, 78,
  \dodoi{10.3847/1538-4357/aa73cf}

\bibitem[{{Vos} {et~al.}(2018){Vos}, {Allers}, {Biller}, {Liu}, {Dupuy},
  {Gallimore}, {Adenuga}, \& {Best}}]{Vos_etal2018}
{Vos}, J.~M., {Allers}, K.~N., {Biller}, B.~A., {et~al.} 2018, \mnras, 474,
  1041, \dodoi{10.1093/mnras/stx2752}

\bibitem[{{Vos} {et~al.}(2022){Vos}, {Faherty}, {Gagn{\'e}}, {Marley},
  {Metchev}, {Gizis}, {Rice}, \& {Cruz}}]{Vos_etal2022}
{Vos}, J.~M., {Faherty}, J.~K., {Gagn{\'e}}, J., {et~al.} 2022, \apj, 924, 68,
  \dodoi{10.3847/1538-4357/ac4502}

\bibitem[{{Vos} {et~al.}(2019){Vos}, {Biller}, {Bonavita}, {Eriksson}, {Liu},
  {Best}, {Metchev}, {Radigan}, {Allers}, {Janson}, {Buenzli}, {Dupuy},
  {Bonnefoy}, {Manjavacas}, {Brandner}, {Crossfield}, {Deacon}, {Henning},
  {Homeier}, {Kopytova}, \& {Schlieder}}]{Vos_etal2019}
{Vos}, J.~M., {Biller}, B.~A., {Bonavita}, M., {et~al.} 2019, \mnras, 483, 480,
  \dodoi{10.1093/mnras/sty3123}

\bibitem[{{Vos} {et~al.}(2020){Vos}, {Biller}, {Allers}, {Faherty}, {Liu},
  {Metchev}, {Eriksson}, {Manjavacas}, {Dupuy}, {Janson}, {Radigan-Hoffman},
  {Crossfield}, {Bonnefoy}, {Best}, {Homeier}, {Schlieder}, {Brandner},
  {Henning}, {Bonavita}, \& {Buenzli}}]{Vos_etal2020}
{Vos}, J.~M., {Biller}, B.~A., {Allers}, K.~N., {et~al.} 2020, \aj, 160, 38,
  \dodoi{10.3847/1538-3881/ab9642}

\bibitem[{{Vos} {et~al.}(2023){Vos}, {Burningham}, {Faherty}, {Alejandro},
  {Gonzales}, {Calamari}, {Bardalez Gagliuffi}, {Visscher}, {Tan}, {Morley},
  {Marley}, {Gemma}, {Whiteford}, {Gaarn}, \& {Park}}]{Vos_etal2023}
{Vos}, J.~M., {Burningham}, B., {Faherty}, J.~K., {et~al.} 2023, \apj, 944,
  138, \dodoi{10.3847/1538-4357/acab58}

\bibitem[{{Werner} {et~al.}(2004){Werner}, {Roellig}, {Low}, {Rieke}, {Rieke},
  {Hoffmann}, {Young}, {Houck}, {Brandl}, {Fazio}, {Hora}, {Gehrz}, {Helou},
  {Soifer}, {Stauffer}, {Keene}, {Eisenhardt}, {Gallagher}, {Gautier}, {Irace},
  {Lawrence}, {Simmons}, {Van Cleve}, {Jura}, {Wright}, \&
  {Cruikshank}}]{Werner_etal2004}
{Werner}, M.~W., {Roellig}, T.~L., {Low}, F.~J., {et~al.} 2004, \apjs, 154, 1,
  \dodoi{10.1086/422992}

\bibitem[{{West} {et~al.}(2009){West}, {Baines}, {Karkoschka}, \&
  {S{\'a}nchez-Lavega}}]{West_etal2009}
{West}, R.~A., {Baines}, K.~H., {Karkoschka}, E., \& {S{\'a}nchez-Lavega}, A.
  2009, in Saturn from Cassini-Huygens, ed. M.~K. {Dougherty}, L.~W.
  {Esposito}, \& S.~M. {Krimigis}, 161, \dodoi{10.1007/978-1-4020-9217-6_7}

\bibitem[{{Zhang} {et~al.}(2013){Zhang}, {West}, {Banfield}, \&
  {Yung}}]{Zhang_etal2013}
{Zhang}, X., {West}, R.~A., {Banfield}, D., \& {Yung}, Y.~L. 2013, \icarus,
  226, 159, \dodoi{10.1016/j.icarus.2013.05.020}

\bibitem[{{Zhou} {et~al.}(2020){Zhou}, {Bowler}, {Morley}, {Apai}, {Kataria},
  {Bryan}, \& {Benneke}}]{Zhou_etal2020}
{Zhou}, Y., {Bowler}, B.~P., {Morley}, C.~V., {et~al.} 2020, \aj, 160, 77,
  \dodoi{10.3847/1538-3881/ab9e04}

\end{thebibliography}
\bibliographystyle{aasjournal}



\end{document}